\newif\ifpdflatex    
\pdflatexfalse         

\documentclass[usenatbib]{mn2e}
\usepackage{graphicx}
\usepackage{amsmath}
\def \s{\hphantom{0}}
\def\lesssim{\mathrel{\hbox{\rlap{\hbox{\lower5pt\hbox{$\sim$}}}\hbox{$<$}}}}
\def\gtrsim{\mathrel{\hbox{\rlap{\hbox{\lower5pt\hbox{$\sim$}}}\hbox{$>$}}}}
\title[Search for Failed Supernovae]{The Search for Failed Supernovae with the Large Binocular Telescope: Constraints from 7 Years of Data}
\author[Adams et al.]
  {\parbox{18cm}{S.~M.~Adams$^{1,2}$, C.~S.~Kochanek$^{2,3}$, J.~R.~Gerke$^{2}$, and K.~Z.~Stanek$^{2,3}$}
  \\
  \\
  $^{1}$ Cahill Center for Astrophysics, California Institute of Technology, Pasadena, CA 91125, USA\\
  $^{2}$ Dept.\ of Astronomy, The Ohio State University, 140 W.\ 18th   Ave., Columbus, OH 43210, USA\\
  $^{3}$ Center for Cosmology and AstroParticle Physics (CCAPP), The Ohio State University, 191 W.\ Woodruff Ave., Columbus, OH 43210, USA\\
  E-mail: sma@astro.caltech.edu}
\begin{document}
\voffset -1.5cm
\maketitle

\begin{abstract}
We report updated results for the first 7 years of our program to monitor 27 galaxies within 10 Mpc using the Large Binocular Telescope to search for failed supernovae (SNe) --- core-collapses of massive stars that form black holes without luminous supernovae.  In the new data, we identify no new compelling candidates and confirm the existing candidate.  Given the 6 successful core-collapse SNe in the sample and one likely failed SN, the implied fraction of core-collapses that result in failed SNe is $f=0.14^{+0.33}_{-0.10}$ at $90\%$ confidence.  If the current candidate is a failed SN, the fraction of failed SN naturally explains the missing high-mass red supergiant SN progenitors and the black hole mass function.  If the current candidate is ultimately rejected, the data implies a $90\%$ confidence upper limit on the failed SN fraction of $f<0.35$.
\\
\\
\end{abstract}


\section{Introduction}

While all stars with $M\gtrsim8~M_{\odot}$ must undergo core collapse, the core-collapse supernova (SN) explosion mechanism is not fully understood.  Despite decades of effort, simulations have difficulty producing robust SN explosions except at the very bottom of this mass range \citep{Janka12,Dolence15,OConnor15,Skinner15,Summa16,Suwa16,Janka16}.  While a majority of core collapses must produce SNe, there is no requirement that all must do so, and it has long been expected that the core collapse of high-mass, low-metallicity stars may fail to explode the star \citep{Heger03}.  More recently, multiple lines of evidence have emerged to suggest that some solar metallicity stars in the local Universe may also result in failed SNe.  

First, there is the deficit of higher mass SN progenitors first identified by \citet{Kochanek08}.  \citet{Smartt09b} showed that the mass distribution of identified SN IIP progenitors truncates $\sim18~M_{\odot}$ even though stellar models predict that massive stars up to $\sim25~M_{\odot}$ should end their lives as red supergiants (RSG) and dubbed this discrepancy the ``red supergiant problem."  This result was confirmed in the study of stellar populations near SN remnants by \citet{Jennings14}.
Even if SN theory cannot yet reliably reproduce the explosion energies of SNe, it can predict the relative ``explodability" of SN progenitors, and studies have consistently shown that the mass range of missing RSG SN progenitors ($18-25~M_{\odot}$) corresponds to progenitor structures that are more difficult to explode \citep{OConnor11,Ugliano12,Pejcha15,Nakamura15,Ertl16,Sukhbold16}.  The existence of these failed SNe would also naturally explain the compact object mass function \citep{Kochanek14b,Kochanek15}.  There is evidence that the massive star formation rate may exceed the SN rate (\citealt{Horiuchi11}, but see \citealt{Botticella12} and \citealt{Xiao15}).  Although long-duration gamma-ray bursts (LGRBs) have been linked to the deaths of Wolf-Rayet stars in Type Ic SNe, there have been three nearby LGRBs for which no associated SN was detected with stringent limits, suggesting that in these cases core-collapse failed to produce or eject nickel (\citealt{Fynbo06,Fryer07,Michalowski16}, but see e.g., \citealt{Yang15}).  Finally we note that Cygnus X-1 \citep{Mirabel03} and the merging $>20~M_{\odot}$ black holes detected by Advanced LIGO \citep{Abbott16a} all likely originally formed from failed SNe \citep{Abbott16b,Belczynski16,Woosley16}.

However, all of this evidence is circumstantial.  Real progress requires observing failed SNe and the properties of their progenitors.
If the shock produced by the core-collapse fails to explode a star, a range of observational signatures may be possible depending on the progenitor mass, structure, and angular momentum.  In some cases the star may simply disappear without any intervening transient \citep{Heger03}.  In others there may be a short (few-day) blue transient powered by fallback accretion onto the newly-formed black hole \citep{Kashiyama15}.  Another possibility for a RSG progenitor, suggested by \citet{Nadezhin80} and numerically verified by \citet{Lovegrove13}, is the expulsion of the weakly-bound hydrogen envelope due to the nearly instantaneous loss of gravitational binding energy from neutrino emission.  This would result in a $3-10$ day, $10^{7}~L_{\odot}$ shock breakout \citep{Piro13} followed by a $\sim$yr, $\sim4000$ K, $\sim10^{6}~L_{\odot}$ transient powered primarily by the hydrogen recombination of the ejected envelope \citep{Lovegrove13}.  

\citet{Kochanek08} proposed and initiated a novel survey monitoring the evolved massive stars in 27 galaxies within 10 Mpc using the Large Binocular Telescope to attempt to directly detect (or at least set meaningful upper limits on) failed SNe arising from RSG progenitors.  Rather than rely on a possible intervening transient to identify failed SNe, the survey is designed to detect the one signature common to all failed SN scenarios: the disappearance of the progenitor star.   Sensitivity to luminosity changes of $10^{4}~L_{\odot}$ in optical filters should be sufficient to detect the disappearance of an evolved $\sim9-30~M_{\odot}$ star \citep[][hereafter referred to as GKS15]{Gerke15}.  
Although ground-based telescopes may not be able to resolve individual stars due to crowding, this sensitivity to differential flux can be achieved with image subtraction.
In practice, the upper mass limit of this survey approach simply represents masses where progenitors are expected to become stripped Wolf-Rayet stars.  The disappearance of many Wolf-Rayet stars cannot be efficiently detected with a broad-band optical survey because their low optical luminosities.

In addition to identifying failed SNe, the survey will produce valuable constraints on SN progenitors (once they fade) and is successfully measuring (or constraining) SN progenitor variability \citep{Szczygiel12,Kochanek17}.
This is the largest survey of its kind and has already been used to constrain the late-time variability of both SN ``impostors" \citep[SN 1997bs;][]{Adams15} and SN 2008S-like events \citep[SN 2008S;][]{Adams16}, identify LBVs \citep{Grammer15}, help characterize dusty stars \citep{Khan15}, and to study the systematic problems in the Cepheid distance scale \citep{Gerke11,Fausnaugh15}.

GKS15 presented constraints on failed SNe from the first 4 years of the survey and identified the first failed SN candidate.  In \citet{Adams16b} we explore this candidate further and find that a failed SN forming a black hole remains the best explanation.  \citet{Reynolds15} also looked for failed SN candidates using archival \emph{HST} data and found one potential candidate.  In this paper we present updated results from this first observational search for failed SNe using the 7 years of data from the LBT survey.
The outline of this paper is as follows: in \S2 we describe the image subtraction and calibration of the data, in \S3 we define the candidate selection, in \S4 we discuss the SNe within the sample, in \S5 we present the candidates, and in \S6 we close with a discussion of the resulting measurement of the fraction of core collapses that result in failed SNe.

\section{Image Subtraction and Calibration}

We largely followed the procedures described in GKS15 for basic data reduction, image subtraction, and calibration.  We continue to use the {\sc isis} image subtraction package \citep{Alard98,Alard00} with the same astrometric reference images used by GKS15, but we created new reference images to be used for our stellar catalogs and image subtraction.  The new reference images are generated from the best $\sim20\%$ of the first 6 years of data, whereas those in GKS15 were generated from the first 3 years of data.  Consequently the new reference images generally have significantly better FWHM and higher S/N, which leads to cleaner image subtractions.  We run the image subtraction pipeline on all epochs regardless of data quality, but only utilize epochs for which the seeing is $<$2\arcsec, the mean background counts are less than 30,000 (to eliminate observations taken during twilight), and the image subtraction scaling factor is greater than 0.4 (to eliminate observations taken through thick cirrus) for candidate selection and in the presented light curves.
The light curves now also include an estimate of the systematic uncertainties based on the RMS photometry of light curves extracted from a grid of points within 10\arcsec of the target source after $3\sigma$-clipping for each epoch.

The procedure for photometric calibrations remains the same as in GKS15.  It is based, when possible, on SDSS stars \citep{Ahn12} with the SDSS $ugriz$ filter system transformed to the $UBVR_{C}$ system according to the prescription in \citet{Jordi06} with Vega magnitudes and zero-points as reported by \citet{Blanton07}.  The handful of exceptions were calibrated as described in GKS15 and the $U$-band data for IC 2574, NGC 925, and NGC 6503 remain uncalibrated.  Updating these remaining calibrations has been on hold pending the release of the PanSTARRS catalogs.

We improved on the masking procedure described in GKS15 by employing separate bad pixel masks to keep track of saturated pixels and the neighboring pixels they could effect during convolutions.  This differs from GKS15 where the images themselves were masked.  This change simplifies the treatment of the regions near saturated stars.  We do not consider photometry from sources within 10 pixels of a saturated pixel or a pixel with no data.  We do, however, track the continuing existence of all saturated sources.

\section{Candidate Selection}
\label{sec:candcrit}

Following GKS15, we generate a master source catalog of ``bright sources" with $\nu L_{\nu}>1000~L_{\odot}$ in the reference image\footnote{This is a lower threshold than in GKS15 because our reference image is constructed from a wider range of epochs and we do not want to lose sources that were $>$$10^{4}~L_{\odot}$ before vanishing from later epochs also used in the reference image} and ``RMS sources" (all sources detected with {\sc SExtractor} in an RMS image generated from the subtracted images for each epoch) for each filter.  We generate {\sc isis} light curves for each of these sources.  
  For a typical galaxy and filter these catalogs include $\sim40,000$ ``bright" sources and $\sim2000$ RMS sources.  Typically there are $\sim300$ sources that are both bright and variable.

We next generate an initial candidate list for each filter using the union of 4 criteria:

1) $\lvert \Delta \nu L_{\nu} \rvert >10^{4}~L_{\odot}$ between all of the following image pairs: first and last, first and penultimate, second and final, and second and penultimate images (where the ``second" image is chosen to be at least one month after the ``first" image).  The multiple image pairs help to eliminate false positives.  For example, requiring a source to be brighter in two exposures eliminates asteroids, while requiring that the first two images are separated by at least a month eliminates most novae from the candidate selection.  Requiring the source to be fainter in the last two images helps to eliminate other false positives such as eclipsing binaries and certain types of subtraction artifacts.
We additionally require the change in flux between the first and last images to be greater than $10\%$, this helps to eliminate luminous variables that are still clearly present in the final image.
Having the benefit of more epochs than GKS15 to help control for variable stars, we allow more flux in the final image for the initial candidate selection.  The motivation being to identify failed SNe that occur within clusters.

2) $\Delta \nu L_{\nu} > 10^5~L_{\odot}$ in at least two consecutive epochs and constrained to exceed this luminosity for 3 months to 3 years.  This filter is intended to select transient events such as those predicted by \citet{Lovegrove13} for a failed SN, as well as luminous core collapse SNe (ccSNe) and stellar mergers, but not novae or asteroids.

3) Saturated at some point, but with $\nu L_{\nu} < 10^{4}~L_{\odot}$ in the last two epochs.  Photometry of saturated sources is not used in the first two candidate selection criteria, but if a source is saturated at some point and is then relatively faint in the final epochs, something clearly faded.  This is intended to select SNe with poor temporal sampling that may cause them to be missed by criteria 2) (e.g., saturated in early epoch, but already faint in the following epoch that, in some cases, may be a full year later) and very bright stars that may be saturated before experiencing a failed SN, while not selecting bright stars fluctuating across the boundary of saturation depending on the seeing.

4) Any of the 235 sources that passed the first round of visual inspection in GKS15 that were not already selected by any of the above criteria.  By revisiting all these sources we can try to recover any failed SN misclassified by GKS15.

We created a master list of the 3314 sources that were selected as candidates under any of these criteria in any filter, and two of the authors (SMA and CSK) independently visually inspected the image subtractions and first/last images of each source in each filter.  The vast majority of sources were either artifacts (2447; poorly subtracted bright stars --- often Galactic foreground stars with a detectable proper motion) or variable stars that exhibited repeated variability and/or were clearly detected in the final epoch.  Only 89 sources survived this initial inspection and were then examined more closely with the aid of the computed light curve, leaving 15 candidates.  Six of these candidates are known SNe we discuss in \S\ref{sec:sne} and the remaining nine candidates for failed SNe are discussed in \S\ref{sec:candidates}.

\begin{table*}
\caption{Galaxy Sample}
\label{tab:galsam}
\begin{tabular}{lcccccc}
\hline \hline
  Galaxy &  Distance &  Number of &  \multicolumn{2}{c}{Observation Period} &  Baseline & Distance \\
         & (Mpc)     & Epochs     & Start & End & (years) & Reference\\
\hline
M81 &  \s3.65 &   39 & 2008-03-08 & 2016-02-09     & 7.8 &  1 \\
M82 &  \s3.52 &    31 & 2008-03-08 & 2016-02-09    & 7.8 & 2 \\
M101 &  \s6.43 &   13 & 2008-03-09 & 2016-02-07    & 7.8 & 3 \\
NGC~628 &  \s8.59 & 21 & 2008-11-25 & 2016-01-04 & 6.8 & 4 \\
NGC~672 &  \s7.20 & 22 & 2008-07-05 & 2015-12-08 & 6.9 & 5 \\
NGC~925 &  \s9.16 & 21 & 2008-07-06 & 2016-01-03 & 7.0 & 6 \\
NGC~2403 & \s3.56 & 38 & 2008-05-05 & 2016-02-11  & 7.1 & 7 \\
NGC~2903 & \s8.90 & 17 & 2008-03-08 & 2016-02-07  & 7.6 & 8 \\
NGC~3077 & \s3.82 & 20 & 2008-05-04 & 2016-02-09  & 7.1 & 5 \\
NGC~3344 & \s6.90 & 18 & 2008-05-04 & 2016-02-07  & 7.1 & 9 \\
NGC~3489 & \s7.18 & 15 & 2008-03-12 & 2016-02-07  & 7.7 & 10 \\
NGC~3627 & 10.62 &  18 & 2008-05-04 & 2016-02-07  & 6.9 & 11 \\ 
NGC~3628 & 10.62 &  18 & 2008-05-04 & 2016-02-07  & 6.9 & 11 \\ 
NGC~4214 & \s2.98 & 14 & 2008-03-13 & 2016-02-07  & 6.9 & 12 \\
NGC~4236 & \s3.65 & 14 & 2008-03-09 & 2016-02-09  & 6.9 & 1 \\
NGC~4248 & \s7.21 & 40 & 2008-03-08 & 2015-05-21  & 6.9 & 13 \\ 
NGC~4258 & \s7.21 & 41 & 2008-03-08 & 2016-01-03  & 7.2 & 13 \\ 
NGC~4395 & \s4.27 & 13 & 2008-03-10 & 2016-02-07  & 4.9 & 14 \\
NGC~4449 & \s3.82 & 16 & 2008-03-09 & 2016-02-09  & 6.9 & 15\\
NGC~4605 & \s5.47 & 15 & 2008-03-13 & 2016-02-09  & 6.9 & 16 \\
NGC~4736 & \s5.08 & 13 & 2008-03-10 & 2016-02-07  & 6.3 & 17 \\
NGC~4826 & \s4.40 & 14 & 2008-03-08 & 2016-02-07  & 7.0 & 2 \\
NGC~5194 & \s8.30 & 21 & 2008-03-09 & 2016-02-07  & 6.3 & 18 \\
NGC~5474 & \s6.43 & 14 & 2008-03-13 & 2016-02-09  & 6.9 & 3 \\
NGC~6503 & \s5.27 & 14 & 2008-05-04 & 2016-02-09  & 6.0 & 6 \\
NGC~6946 & \s5.96 & 38 & 2008-05-03 & 2015-12-08  & 7.4 & 19 \\
IC~2574 & \s4.02 &  18 & 2008-03-13 & 2016-02-09  & 7.7 & 6 \\
\hline
\multicolumn{7}{p{0.8\textwidth}}{The baseline is the time from the second
observation to the penultimate observation in the selection period.  
References -- (1) \citet{Gerke11};
(2) \citet{Jacobs09}; (3) \citet{Shappee11}; (4) \citet{Herrmann08};
(5) \citet{Karachentsev04}; (6) \citet{Karachentsev03}; (7) \citet{Willick97};
(8) \citet{Drozdovsky00}; (9) \citet{Verdes00}; (10) \citet{Theureau07};
(11) \citet{Kanbur03}; (12) \citet{Dopita10};
(13) \citet{Herrnstein99}; (14) \citet{Thim04};
(15) \citet{Annibali08}; (16) \citet{Karachentsev06};
(17) \citet{Tonry01}; (18) \citet{Poznanski09};
(19) \citet{Karachentsev00}.
}
\end{tabular}
\end{table*}

\section{Supernovae}
\label{sec:sne}

During the survey period analyzed (see Table \ref{tab:galsam}), 6 core-collapse SNe (SN 2009hd, SN 2011dh, SN 2012fh, SN 2013ej, SN 2013em, and SN 2014bc) and 2 SNe Ia (SN 2011fe and SN 2014J) occurred in our galaxy sample.  
SN 2009hd \citep{Monard09, EliasRosa11} and SN 2011dh \citep{Griga11, Szczygiel12c} were already discussed in GKS15.

SN 2012fh, a Type Ic SN in NGC 3344 \citep{Nakano12}, is not selected as a candidate because the SN had already faded below $10^{5}~L_{\odot}$ in each filter by the first post-peak epoch on 6 June 2013 and the progenitor flux was below $10^{4}~L_{\odot}$ in all four filters.  
SN 2013am, a low-velocity Type IIP in NGC 3623 \citep{Nakano13,Zhang14}, was selected as a candidate by criteria 2) and 3), but has not yet faded below the progenitor luminosity.  No luminosity constraints on the progenitor have been published.
Likewise, SN 2013ej, a Type IIP in NGC 628 with signs of weak interaction \citep{Kim13,Bose15}, was selected as a candidate by criterion 2), but has not yet faded below the progenitor luminosity.  Archival \emph{HST} imaging constrains the progenitor to be an M-type supergiant with $8-15.5~M_{\odot}$ \citep{Fraser14} and modeling of the SN light curve also suggests a RSG progenitor of $12-13~M_{\odot}$ \citep{Huang15}.
The final SN to occur within the candidate selection window was SN 2014bc, a SN IIP in NGC 4258 \citep{Smartt14}.  Being very close to the core of the galaxy, the source location is saturated in the $B$, $V$, and $R_{C}$ bands.  The SN was selected as a ``burst" candidate (criteria 2) in $U$-band, but is still brighter than the progenitor in this filter.  

Of the 6 core-collapse SNe, only SN 2011dh would be selected as a failed SN candidate without a detection of the transient.  For SN 2009hd, SN 2013am, and SN 2013ej the problem is simply that the SN has not yet faded significantly below the progenitor flux.  
SN 2014bc would not have been selected because the only bands in which the disappearance of the progenitor flux normally would have been detected are saturated at the SN's location near the host galaxy's core.  
The core collapse of SN 2012fh probably would not be identified in the absence of the SN because the stripped core progenitor was likely optically faint due to the large bolometric correction from the high temperature of the progenitor
(see Fig. 1 in GKS15).

Additionally, SN 2008S, which might have been a low-energy SN \citep{Prieto08,Adams16}, was caught during its outburst and selected as a candidate.  We do not include in our analysis because the transient began a few months before the start of the survey.
The number of SNe that occurred in our sample galaxies during the survey and our selection efficiency for them will be used in \S\ref{sec:rate} to compute constraints on the rate of failed SNe.


\section{Candidates}
\label{sec:candidates}

\begin{table*}
\caption{Candidate List}
\label{tab:candidates}
\begin{tabular}{lccccccc}
\hline \hline
	&	&	& Candidate & $L_{R,\mathrm{i}}-L_{R,\mathrm{f}}$ & $L_{R,\mathrm{max}}-L_{R,\mathrm{min}}$ & $t_{\mathrm{f}}$ & 		\\
ID	& RA	& Dec	& Criteria  & [$L_{\odot}$]			  & [$L_{\odot}$]			    & [days]	       & Classification \\
\hline
SN 2008S  & 20:34:45.36 & +60:05:58.2 & 1        & $\hphantom{-}5.2\times10^{5}$ & $\hphantom{>}5.2\times10^{5}$ & 326--535 & SN IIn \\
SN 2011dh & 13:30:05.15 & +47:10:11.8 & 1,2      & $\hphantom{-}5.5\times10^{4}$ & $>6.5\times10^{6}$ & 328--622 & SN IIb \\
SN 2011fe & 14:03:05.75 & +54:16:25.3 & 2,3      & $\hphantom{-}3.1\times10^{3}$ & $>1.9\times10^{6}$ & 82--554 & SN Ia \\
SN 2013am & 11:18:56.93 & +13:03:45.0 & 2,3      & $-5.1\times10^{3}$ & $>1.4\times10^{7}$ & 0--300 & SN IIP \\
SN 2013ej & 01:36:48.19 & +15:45:30.7 & 2        & $-1.1\times10^{4}$ & $>3.7\times10^{6}$ & 69--680 & SN IIP \\ 
SN 2014bc & 12:18:57.72 & +47:18:11.2 & 1,3      & $\hphantom{-}9.2\times10^{3}$ & $>5.0\times10^{6}$ & 0--171 & SN IIP \\ 
N6946-BH1 & 20:35:27.56 & +60:08:08.3 & 1,2      & $\hphantom{-}6.0\times10^{4}$ & $\hphantom{>}7.8\times10^{5}$ & 89--471 & FSN \\
PSN J14021678 \\
$\hphantom{...}$  +5426205 & 14:02:16.80 & +54:26:20.7 & 1,2,3   & $\hphantom{-}9.2\times10^{4}$ & $>1.6\times10^{6}$ & 122--550 & merger \\
N2903-SF1 & 09:32:11.90 & +21:29:00.5 & 1	& $\hphantom{-}1.3\times10^{5}$ & $\hphantom{>}1.4\times10^{5}$ & 1805--2390 & SF \\
N2903-SF2 & 09:32:08.84 & +21:31:36.2 & 1	& $\hphantom{-}2.3\times10^{4}$ & $\hphantom{>}2.5\times10^{4}$ & 1710--1768 & SF \\
N5194-SF1 & 13:29:50.60 & +47:10:50.6 & 1	& $\hphantom{-}1.1\times10^{5}$ & $\hphantom{>}1.1\times10^{5}$ & 1806--2238 & SF \\
N6503-SF1 & 17:49:32.66 & +70:08:09.0 & 1	& $\hphantom{-}2.0\times10^{4}$ & $\hphantom{>}5.9\times10^{4}$ & 1440--1686 & SF \\ 
N6946-SF1 & 20:34:14.29 & +60:03:01.1 & 1	& $\hphantom{-}1.1\times10^{4}$ & $\hphantom{>}1.1\times10^{4}$ & 2225--2244 & SF \\
N6946-SF2 & 20:35:11.32 & +60:08:49.2 & 1	& $\hphantom{-}8.7\times10^{4}$ & $\hphantom{>}1.1\times10^{4}$ & 947--954 & SF \\ 
N925-OC1  & 02:27:21.88 & +33:34:05.0 & 1	& $\hphantom{-}1.5\times10^{4}$ & $\hphantom{>}1.9\times10^{4}$ & 359--715 & OC \\
\hline
\multicolumn{8}{p{0.99\textwidth}}{List of candidates that passed the final round of visual inspection.  `Candidate criteria' are those listed in \S\ref{sec:candcrit}.  $L_{R,\mathrm{i}}$ is the $R$ band luminosity of the first epoch and $L_{R,\mathrm{f}}$ is the $R$ band luminosity of the final epoch.  $L_{R,\mathrm{max}}$ and $L_{R,\mathrm{min}}$ are the maximum and minimum $R$ band luminosities observed for each source in the LBT lightcurves. $t_{\mathrm{f}}$ is the constraint on the time for the source flux to decrease from $L_{R,\mathrm{max}}$ by $0.9(L_{R,\mathrm{max}}-L_{R,\mathrm{min}})$. FSN=failed SN, SF=slowly fading candidate ($t_{\mathrm{f}}>730$ days), and OC=other candidate.  N6946-BH1 was previously identified as a failed SN candidate in GKS15 and is the thoroughly discussed in \citet{Adams16b}.  The SF candidates fade too slowly to be failed SNe (see \S\ref{sec:sf}).  Given the limited data, N925-OC1 is consistent with a failed SN, but requires further vetting (see \S\ref{sec:oc}).}
\end{tabular}
\end{table*}

Excluding these known SNe, nine candidates passed the second round of visual
inspection (see Table \ref{tab:candidates}). Among these nine candidates is the failed SN candidate already identified
in GKS15 (which we hereafter will refer to as N6946-BH1) and the
likely stellar merger PSN J14021678+5426205 \citep{Goranskij16,Blagorodnova16}. There are also six sources that fade
too slowly to be promising failed SN candidates and one source that requires further
observations to be deemed a promising candidate. We will discuss each of these
sources below, but ultimately consider N6946-BH1 to be the only likely failed SN in
the first 7 years of the survey.

%
%

\subsection{N6946-BH1}
\label{sec:N6946-BH1}

The most promising candidate was previously announced in GKS15 and it is discussed in-depth in \citet{Adams16b}.  Here we will only briefly summarize the candidate.  It is located in NGC 6946 at RA 20:35:27.56 and Dec +60:08:08.29.  The progenitor was a $\sim25~M_{\odot}$ RSG which can be traced back to 1999 in archival data from other sources.  After a $\sim10^{6}~L_{\odot}$ outburst in 2009, it disappeared from the optical within a few months.  It faded more slowly in the IR, but the bolometric luminosity now appears to be significantly fainter than the progenitor.  In \citet{Adams16b}, we support the conclusion of GSKS15 that this source is a promising failed SN candidate.  We present new \emph{HST} imaging confirming the optical disappearance of the progenitor and find that a surviving star cannot be hidden by simple models of dust formation in material from a mass loss episode or wind.  We propose that the long IR transient may be fallback accretion onto a newly formed black hole obscured by dust that has formed in the ejected envelope of the progenitor.  An X-ray detection would confirm this candidate as a failed SN, but the weakly-ejected envelope may prevent such a detection.  Ultimately, observations with the \emph{James Webb Space Telescope} may be needed to confirm that a surviving star is not hidden by cold dust.

\subsection{PSN J14021678+5426205}

\begin{figure}
  \includegraphics[width=8.6cm, angle=0]{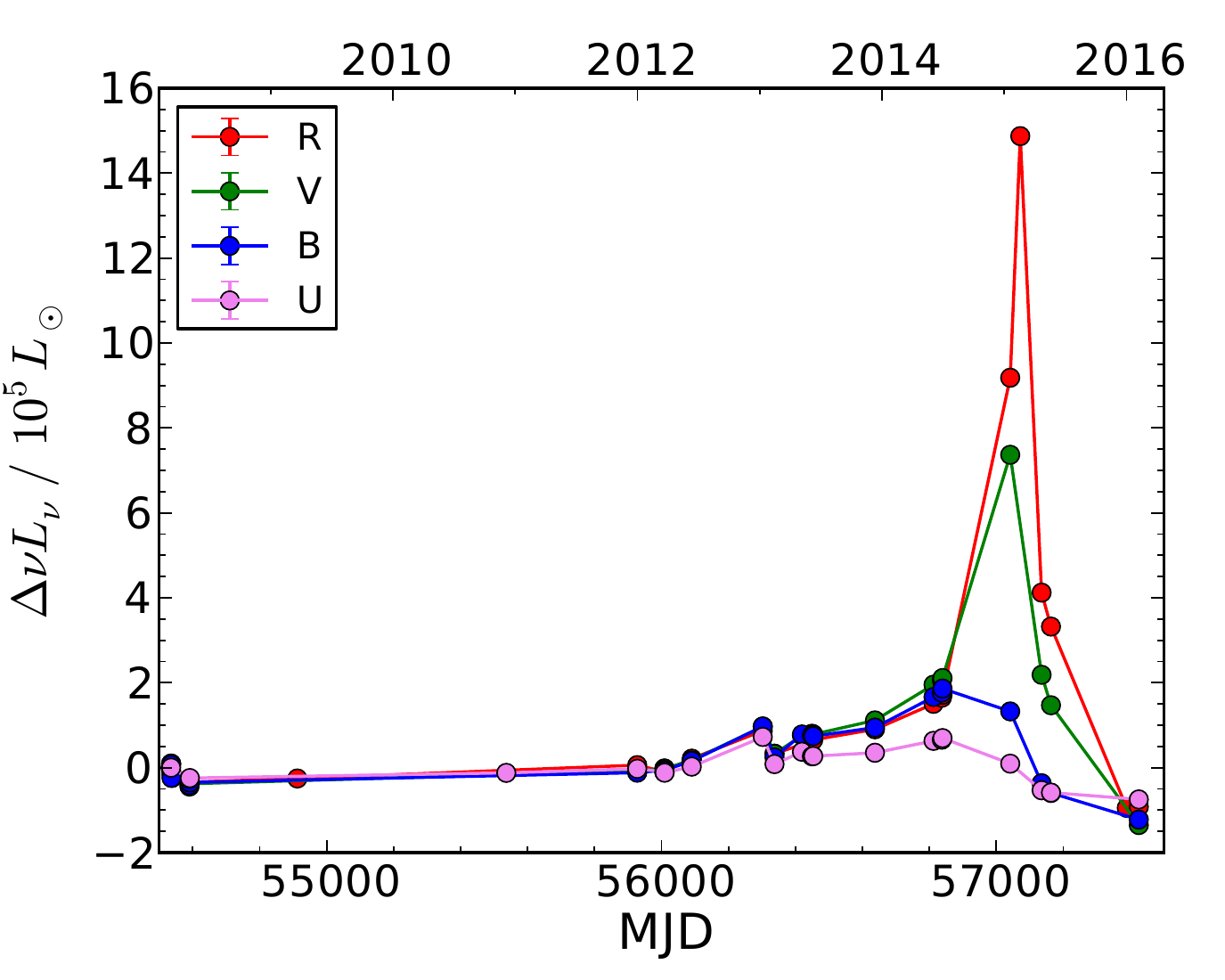}
  \caption{Differential (relative to the first epoch) light curves of PSN J14021678+5426205.  When not shown, statistical uncertainties are smaller than the size of the points. \label{fig:PSNlc}}
\end{figure}

\begin{figure*}
  \includegraphics[width=0.99\textwidth]{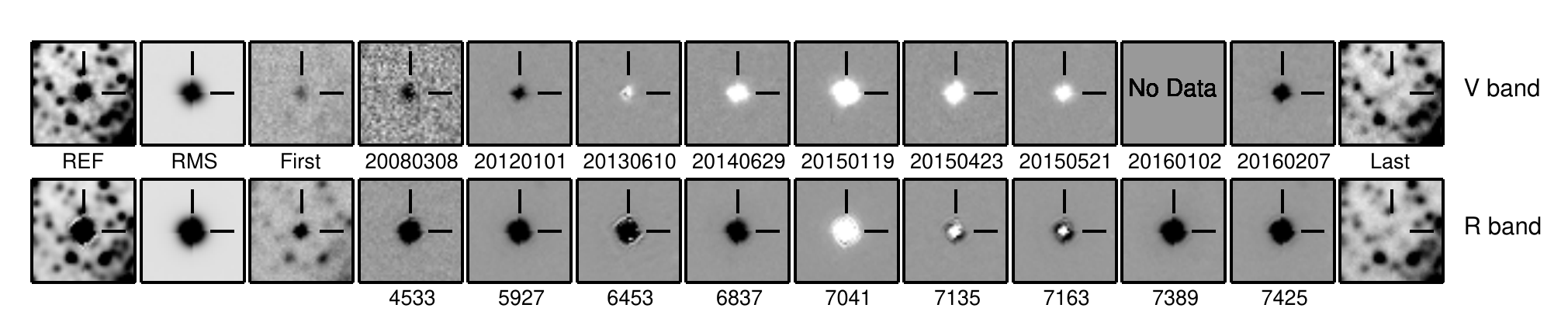}
  \caption{$V$ and $B$ band imaging for PSN J14021678+5426205.  In the reference (REF), RMS, First, and Last images the darker the source, the brighter it is.  The middle columns labeled with dates (top) and the last 4 digits of the Julian Date (bottom) are subtracted images for selected epochs --- individual epoch minus the REF --- in which darker shades mean the source is dimmer than in the reference image and whiter shades mean it is brighter.  The transient is saturated in the $V$ and $R_{\mathrm{C}}$ images from 23 April 2015 and 21 May 2015. \label{fig:PSNstamps}}
\end{figure*}

One of the sources identified as a candidate is PSN J14021678+5426205, which is believed to have been a stellar merger in M101 \citep{Goranskij16,Blagorodnova16}.  PSN J14021678+5426205 satisfied all 3 of the criteria for new candidates ($\Delta \nu L_{\nu} < -10^{4}~L_{\odot}$, $\nu L_{\nu}>10^{5}~L_{\odot}$ for between 3 months and 3 years, and saturated at some point but $\nu L_{\nu}<10^{4}~L_{\odot}$ in the final two epochs; see Fig. \ref{fig:PSNlc} and \ref{fig:PSNstamps}).

The survey data showed that a luminous ($10^{5.3}~L_{\odot}$, 7500 K) progenitor was variable prior to mid-2012 and then steadily brightened.  In the final pre-transient epoch it brightened dramatically in $V$ and $R$ but faded in $U$ \citep{Gerke15b}.  After a $>10^{6}~L_{\odot}$ eruption the progenitor disappeared from the optical but became a very bright mid-IR source \citep{Blagorodnova16}.  At late-times the IR flux still exceeds the bolometric luminosity of the progenitor.  The slow luminosity rise and cooling photosphere, followed by a cool optical transient, profuse dust formation, and a long IR transient are all characteristic of a stellar merger \citep{Crause03,Pejcha16,Pejcha16b,MacLeod16}.

Although it is clear that PSN J14021678+5426205 is not a failed SN, it does raise the question of whether other, fainter stellar mergers missed by SN surveys (and thus without spectroscopy and well-sampled light curves) may be identified as a failed SN candidate.  Without data beyond the LBT survey, PSN J14021678+5426205 would remain a possible failed SN candidate.  This illustrates the importance of acquiring late-time mid and near-IR photometry for any failed SN candidate (as we do for N6946-BH1 in \citealt{Adams16b}).

How many mergers might we expect in our survey?  
Using the empirical relation found by \citet{Kochanek14d} between stellar mass and merger luminosity, a stellar merger involving an $8~M_{\odot}$ ($25~M_{\odot}$) RSG progenitor would have a peak magnitude of $M_{V}\sim-10.1$ ($M_{V}\sim-13.7$).  This corresponds to apparent magnitudes of $m_{V}\sim17.3-20.0$ ($m_{V}\sim13.7-16.4$) at the distances of the galaxies in our sample.  
Given that the sample galaxies are popular targets of amateur SN hunters and have also been monitored with professional SN surveys (CRTS, PTF, ASAS-SN, and MASTER) for the majority of the survey period, we would expect nearly all mergers involving massive stars in the closer galaxies to be discovered, as well as a majority in the most distant galaxies in the sample.  Scaling from \citet{Kochanek14d} we would expect 
$3.4^{+8.6}_{-2.6}$ mergers per century in a Milky Way-like galaxy.  Assuming stellar mass is proportional the absolute $B$-band magnitude of a galaxy and adopting the $B$-band magnitude of the Andromeda Galaxy for that of the Milky Way, the typical galaxy in the LBT sample has a merger rate of $\sim1/3$ the Galactic rate.  Thus, we would expect 
$2.1^{+5.4}_{-1.6}$ 
mergers with peak luminosities of $M_{V}\le-10$ in the survey, consistent with PSN J14021678+5426205 being the only likely candidate for a merger in the data.



\subsection{Slowly Declining Sources}
\label{sec:sf}

Six candidates disappear in the optical after slowly fading over the course of
several years. We briefly present each one and then discuss why these slowly fading
candidates are unlikely to be failed SNe.

\begin{figure}
  \includegraphics[width=8.6cm, angle=0]{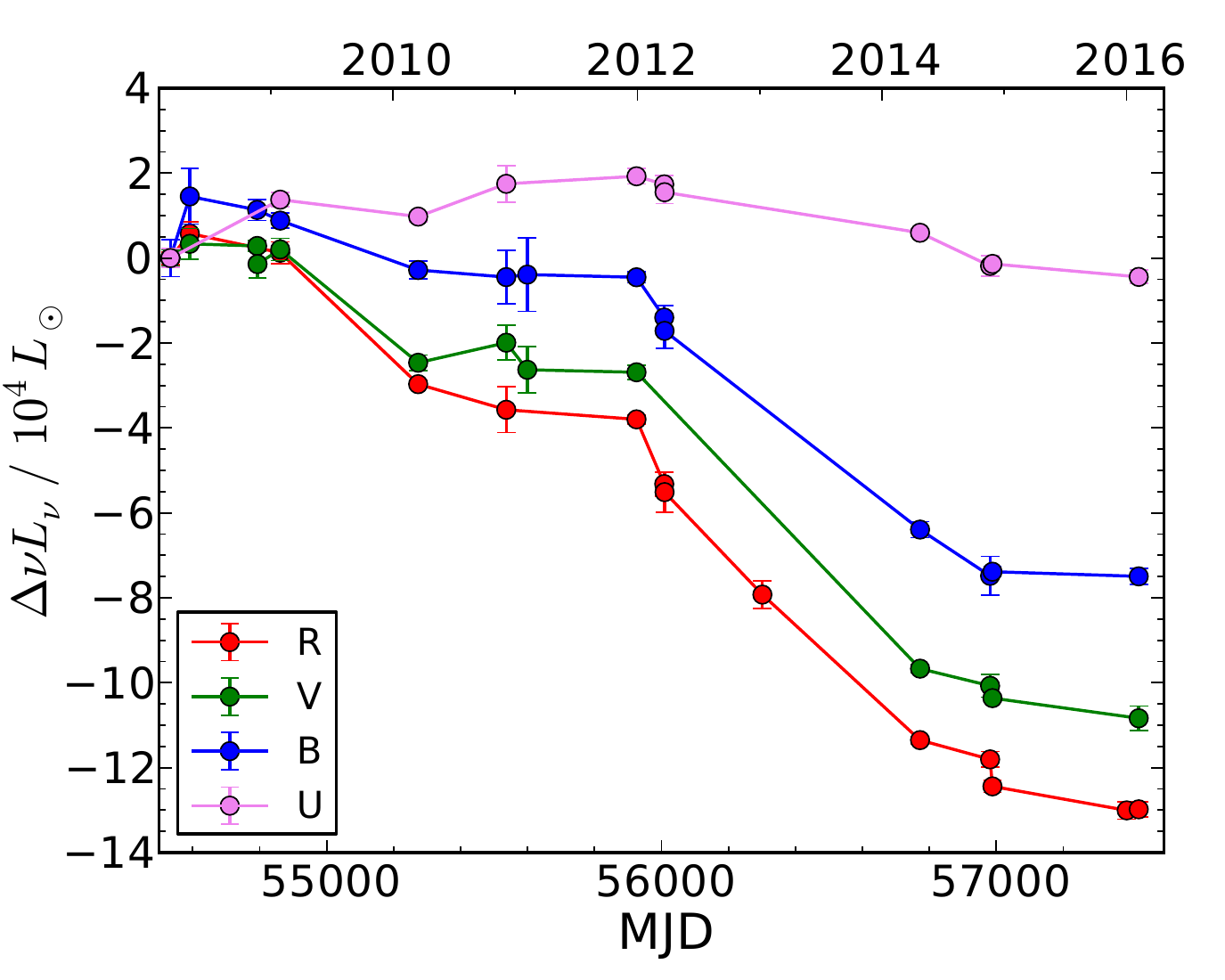}
  \caption{Same as Fig. \ref{fig:PSNlc}, but for N2903-SF1. \label{fig:N2903-SD1lc}}
\end{figure}

\begin{figure*}
\begin{center}
  \includegraphics[width=0.99\textwidth]{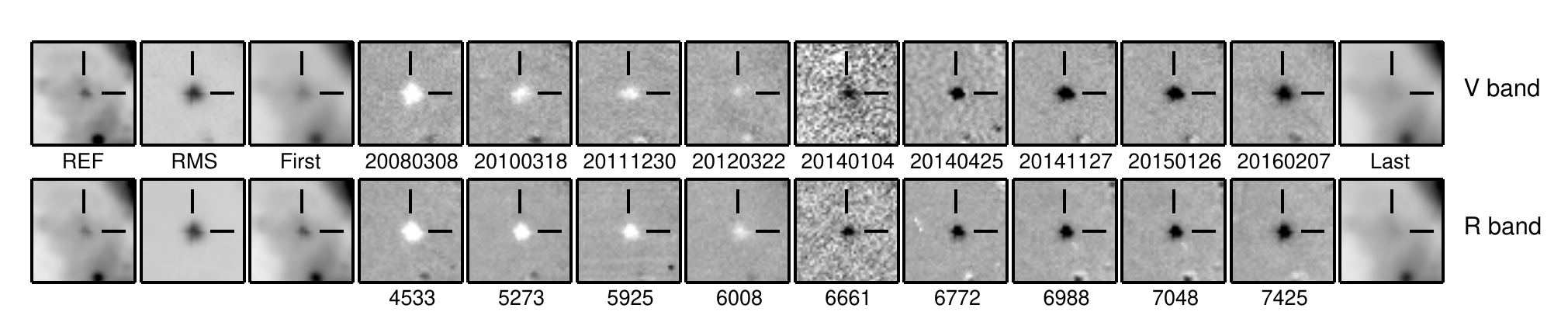}
  \caption{Same as Fig. \ref{fig:PSNstamps}, but for N2903-SF1.\label{fig:N2903-SD1stamps}}
\end{center}
\end{figure*}

$\bullet$ N2903-SF1 is in NGC 2903 at RA 09:32:11.90 and Dec +21:29:00.5.  This bright source in a crowded region faded by $\sim10^{5}~L_{\odot}$ roughly continuously over the survey period (Fig. \ref{fig:N2903-SD1lc}).  Although no obvious source remains in the final LBT epoch (Fig. \ref{fig:N2903-SD1stamps}), a source is still resolved in an epoch on 27 November 2014 that had exceptionally good seeing.  Archival \emph{HST} imaging from 1994 (PI: J. Trauger, GTO-5211) and 2004 (PI: L. Ho, SNAP-9788) shows an isolated point-source coincident with the LBT source.

\begin{figure}
\begin{center}
  \includegraphics[width=8.6cm, angle=0]{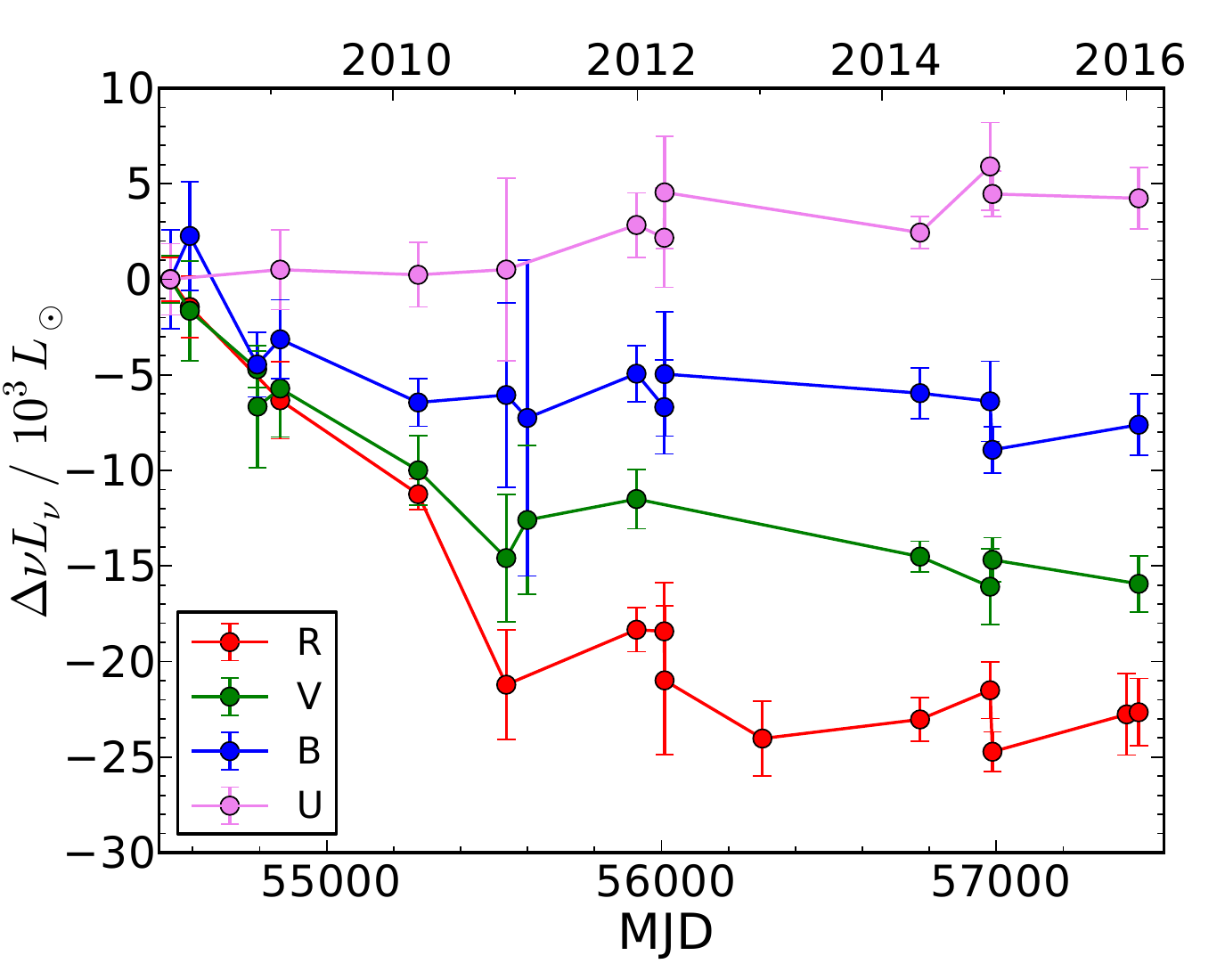}
  \caption{Same as Fig. \ref{fig:PSNlc}, but for N2903-SF2. \label{fig:N2903-SD2lc}}
\end{center}
\end{figure}

\begin{figure*}
\begin{center}
  \includegraphics[width=0.99\textwidth]{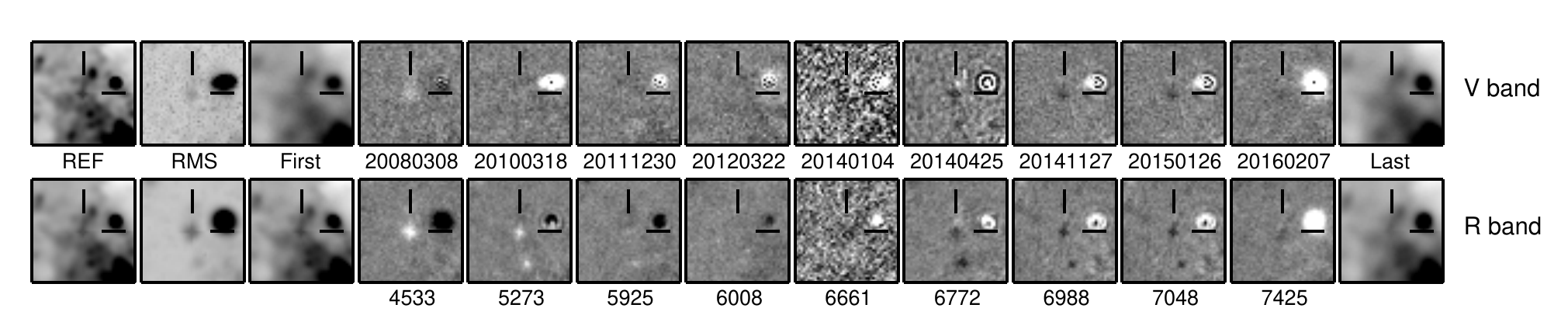}
  \caption{Same as Fig. \ref{fig:PSNstamps}, but for N2903-SF2.\label{fig:N2903-SD2stamps}}
\end{center}
\end{figure*}

$\bullet$ Another source in NGC 2903 located at RA 9:32:08.84 and Dec +21:31:36.2, N2903-SF2, faded $\sim2\times10^{4}~L_{\odot}$ over the first 3 years of the survey (Fig. \ref{fig:N2903-SD2lc}).  There is still flux at the source location in the final LBT epoch but it is unclear whether it is associated with a surviving star since the progenitor was located in a crowded field (Fig. \ref{fig:N2903-SD2stamps}).  The likely progenitor is easily resolved in archival \emph{HST} imaging from 2001 (PI: M. Regan, SNAP-8597) and 2004 (PI: L. Ho, SNAP-9788).

\begin{figure}
\begin{center}
  \includegraphics[width=8.6cm, angle=0]{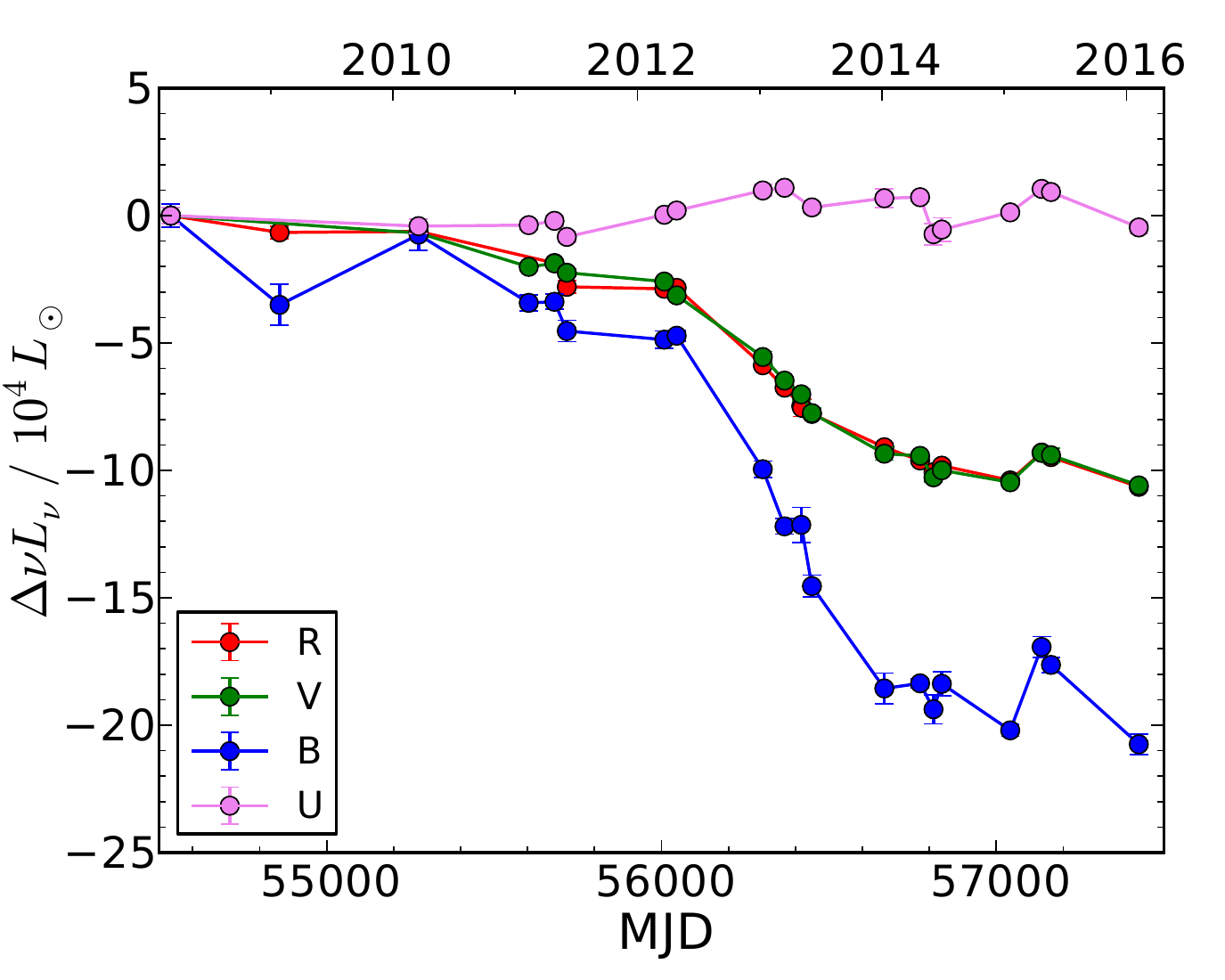}
  \caption{Same as Fig. \ref{fig:PSNlc}, but for N5194-SF1. \label{fig:N5194-SD1lc}}
\end{center}
\end{figure}

\begin{figure*}
\begin{center}
  \includegraphics[width=0.99\textwidth]{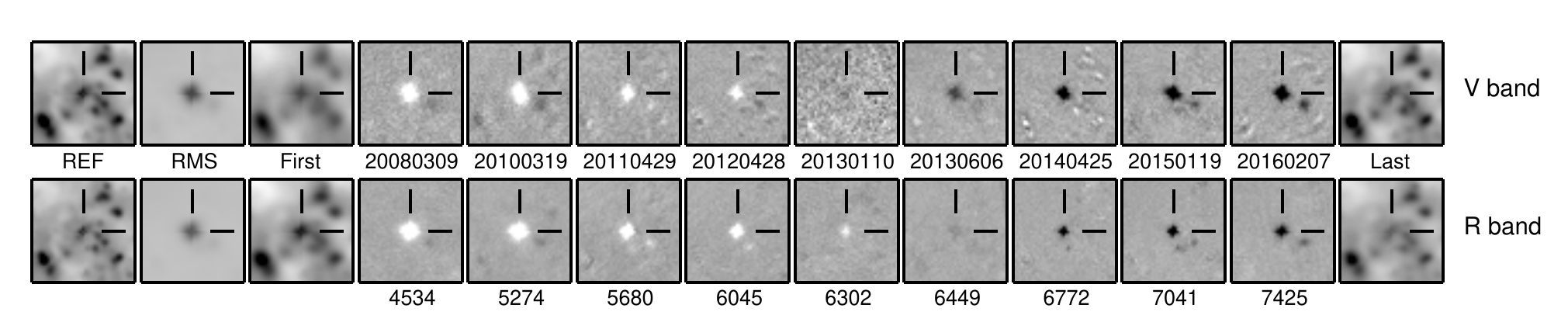}
  \caption{Same as Fig. \ref{fig:PSNstamps}, but for N5194-SF1.\label{fig:N5194-SD1stamps}}
\end{center}
\end{figure*}

$\bullet$ N5194-SF1 is in NGC 5194 at RA 13:29:50.60 and Dec +47:10:50.6.  This source faded in $RVB$ by $>$$10^{5}~L_{\odot}$ over four years (Fig. \ref{fig:N5194-SD1lc}).  There is also flux at this source's location in the final LBT epoch, but it was not eliminated in the visual inspection because it is unclear whether the flux is from a surviving star or the cluster in which the progenitor resided (Fig. \ref{fig:N5194-SD1stamps}).  Archival \emph{HST} $F275$ and $F336$ imaging from 11 September 2014 (PI: S. van Dyk, GO-13340) reveals a bright UV source at the source location.

\begin{figure}
  \includegraphics[width=8.6cm, angle=0]{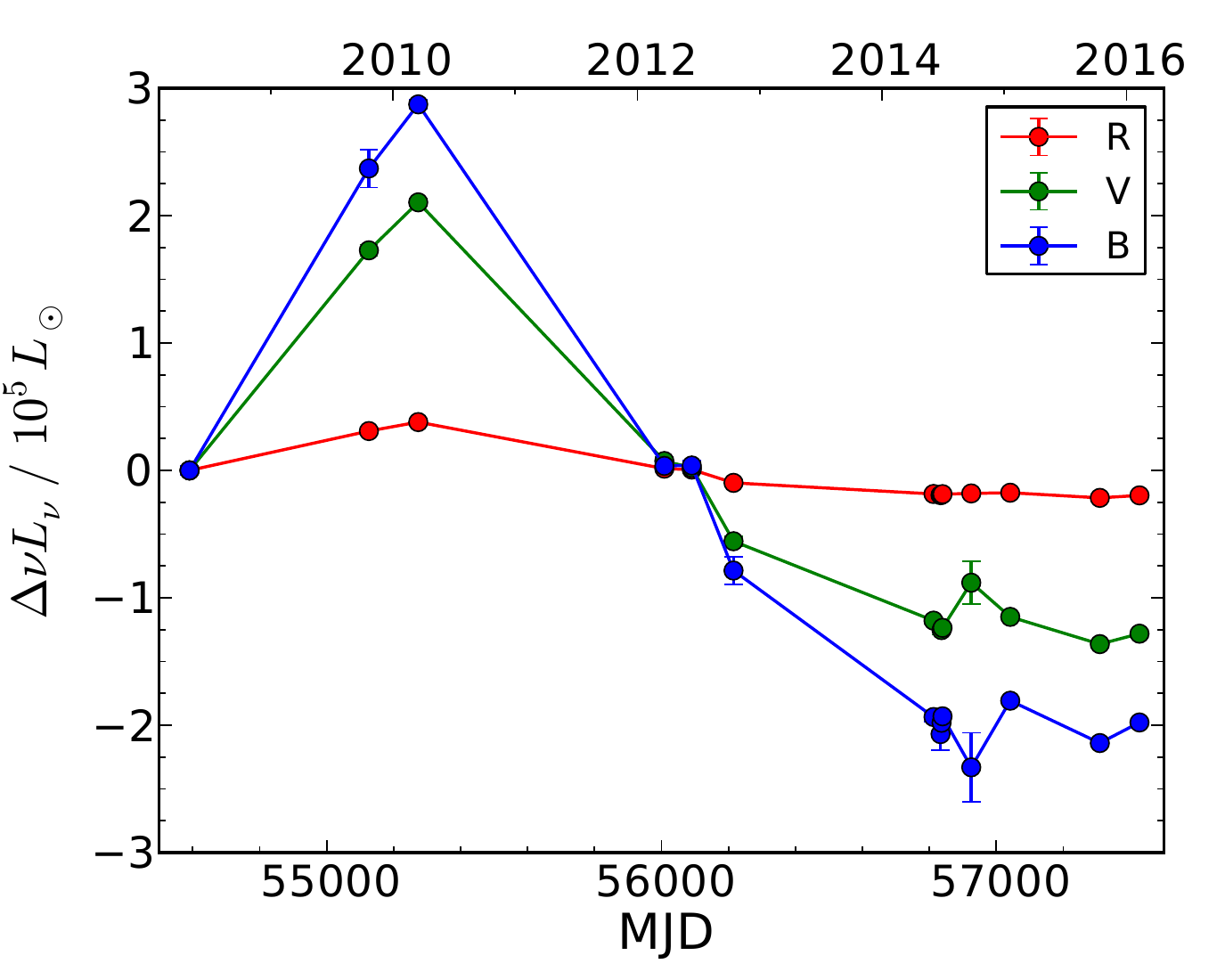}
  \caption{Same as Fig. \ref{fig:PSNlc}, but for N6503-SF1. \label{fig:N6503-C1lc}}
\end{figure}

\begin{figure*}
\begin{center}
  \includegraphics[width=0.99\textwidth]{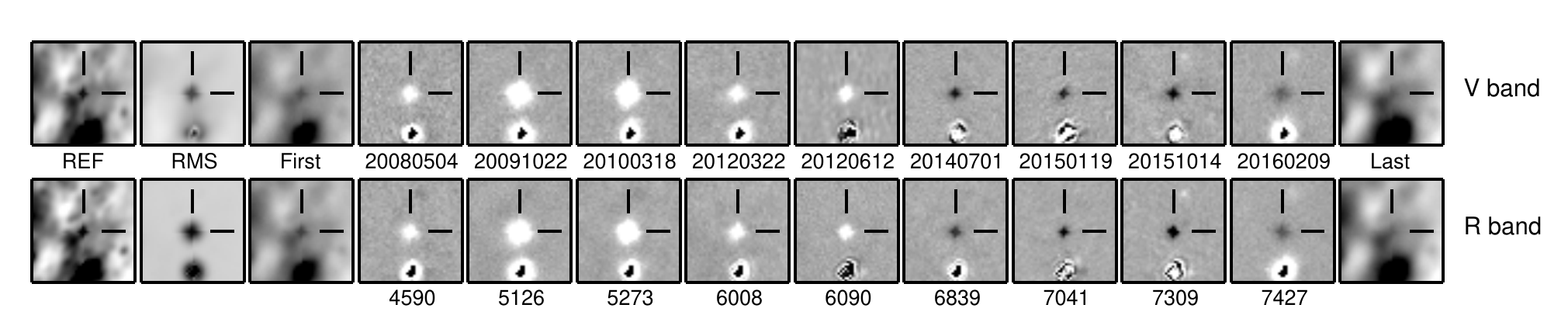}
  \caption{Same as Fig. \ref{fig:PSNstamps}, but for N6503-SF1.\label{fig:N6503-C1stamps}}
\end{center}
\end{figure*}

$\bullet$ N6503-SF1 is in NGC 6503 at RA 17:49:32.66 and Dec +70:08:09.0.  This source produced a $>$$10^{5}~L_{\odot}$ transient that lasted at least 5 months and possibly longer than 6 years (depending on whether the first epoch was the quiescent source or part of the outburst; Fig. \ref{fig:N6503-C1lc}).  The transient is much hotter than expected for a stellar merger.  In the final LBT epoch there is likely a remaining source, but given the crowded nature of the environment this is not conclusively a single surviving star (Fig. \ref{fig:N6503-C1stamps}).  However, in archival \emph{HST} data from 21 August 2013 (PI: D. Calzetti, GO-13364) there is clearly a bright point source remaining.

\begin{figure}
  \includegraphics[width=8.6cm, angle=0]{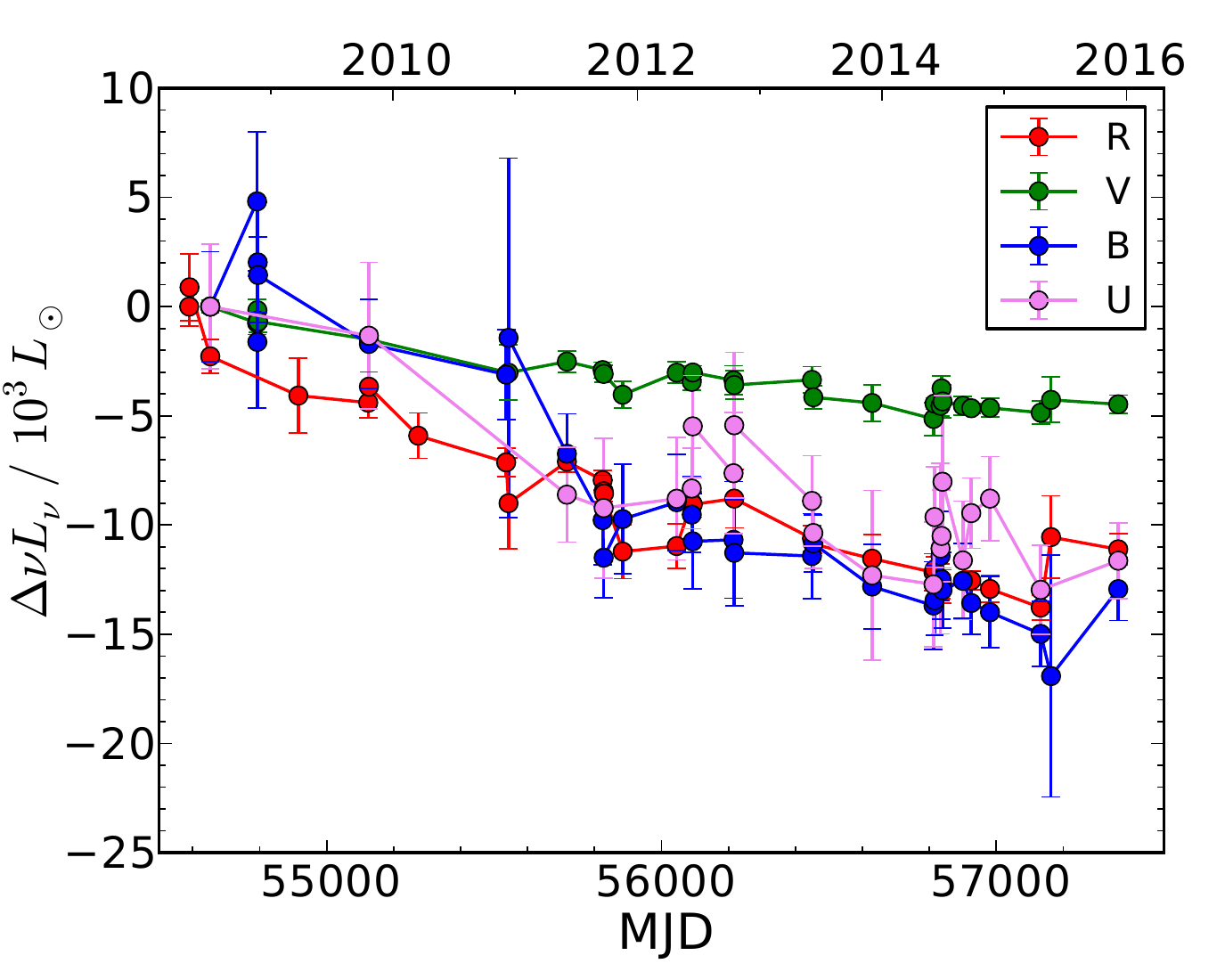}
  \caption{Same as Fig. \ref{fig:PSNlc}, but for N6946-SF1. \label{fig:N6946-C1lc}}
\end{figure}

\begin{figure*}
\begin{center}
  \includegraphics[width=0.99\textwidth]{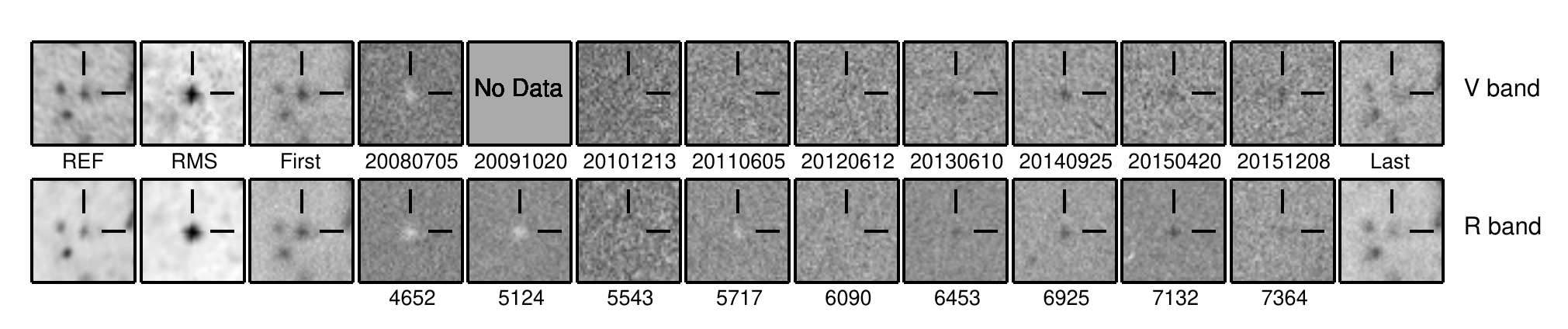}
  \caption{Same as Fig. \ref{fig:PSNstamps}, but for N6946-SF1.\label{fig:N6946-C1stamps}}
\end{center}
\end{figure*}

$\bullet$ N6946-SF1 is located at RA 20:34:14.29 and Dec +60:03:01.1 and it faded over the entire duration of the survey from an initial luminosity of $\sim2\times 10^{4}~L_{\odot}$ to $<10^{4}~L_{\odot}$ (Fig. \ref{fig:N6946-C1lc}).  The $R$ band flux increased modestly above its minimum for the final two epochs and a source is still detected in the final $R$ band image (Fig. \ref{fig:N6946-C1stamps}). 

\begin{figure}
  \includegraphics[width=8.6cm, angle=0]{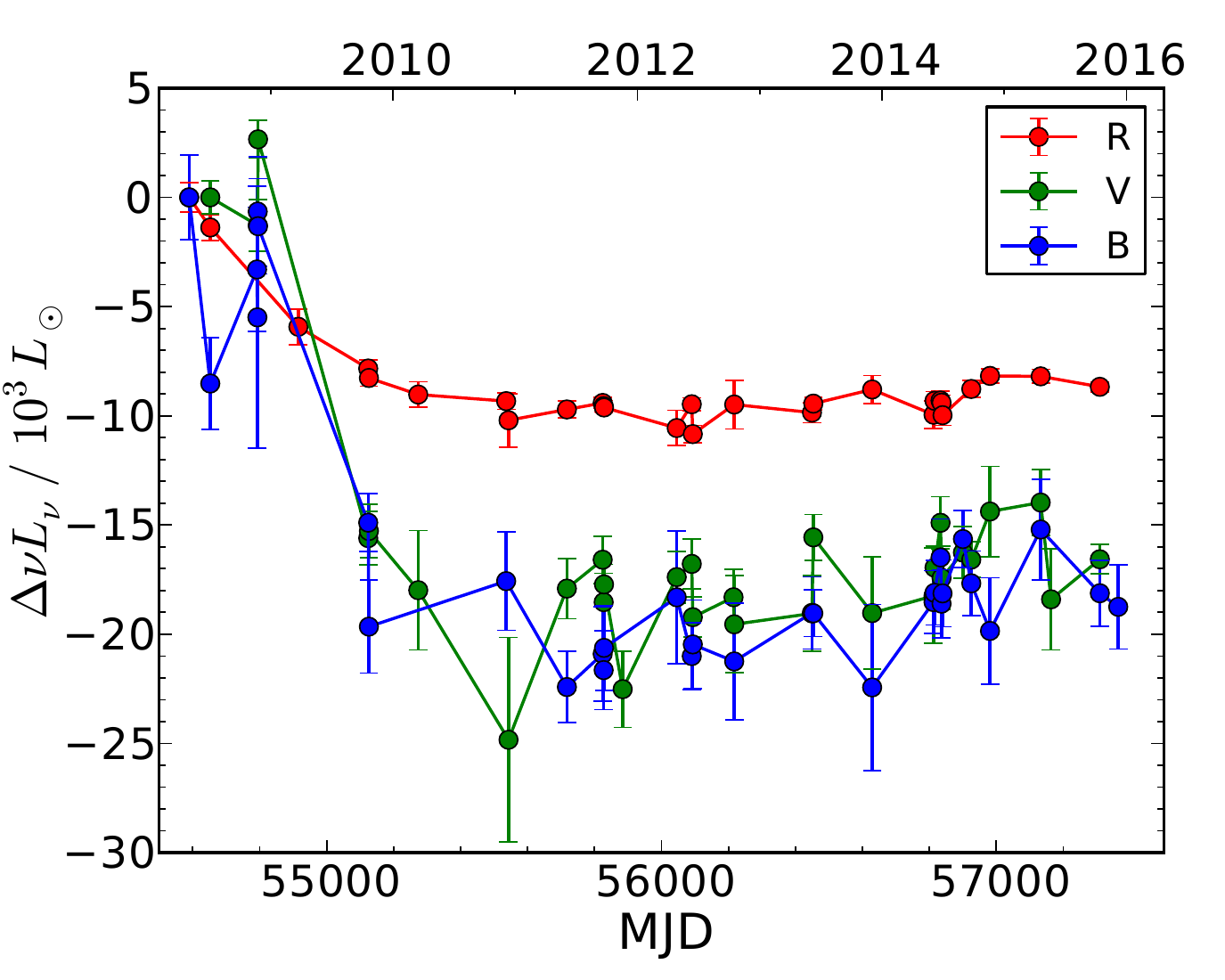}
  \caption{Same as Fig. \ref{fig:PSNlc}, but for N6946-SF2. \label{fig:N6946-C2lc}}
\end{figure}

\begin{figure*}
\begin{center}
  \includegraphics[width=0.99\textwidth]{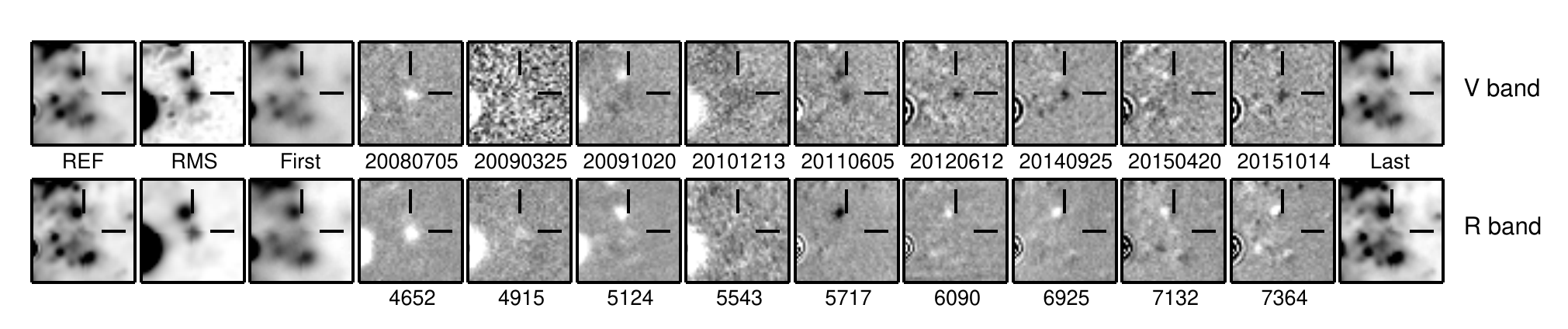}
  \caption{Same as Fig. \ref{fig:PSNstamps}, but for N6946-SF2.\label{fig:N6946-C2stamps}}
\end{center}
\end{figure*}

$\bullet$ Another source in NGC 6946 located at RA 20:35:11.32 and Dec +60:08:49.2, N6946-SF2, faded by $\sim$$10^{4}~L_{\odot}$ over the first 2 years of survey before remaining roughly constant for the last 6 years (Fig. \ref{fig:N6946-C2lc}).  There appears to be a remaining source in the final LBT epoch, although the field is very crowded (Fig. \ref{fig:N6946-C2stamps}).  There is also a bright near-IR source in \emph{HST} data from 9 February 2016 (A. Leroy, GO-14156).

Although there is recent \emph{HST} data for some of these candidates that show a surviving star, in other cases there is no \emph{HST} data and it is difficult to rule out the disappearance of a star using only ground-based imaging.  
Still, the long times scales provide a physical reason to reject these sources.  In the failed SN model of \citet{Lovegrove13} the loss of gravitational binding energy from the neutrino emission at the time of core-collapse can result in the ejection of the weakly-bound hydrogen envelope of a RSG.  Without radioactive ejecta, a transient associated with the ejecta can only be powered by recombination, leading to strict limits on the luminosity and lifetime of any associated transient.
\citet{Popov93} showed that the duration of the recombination-powered plateau phase, $t_{\mathrm{p}}$, of a ccSN is
\begin{eqnarray}
t_{\mathrm{p}}\approx 99 \left(\frac{\kappa}{0.34~\mathrm{cm}^{2}\mathrm{g}^{-1}}\right)^{1/6} \left(\frac{M_{\mathrm{ej}}}{10~M_{\odot}}\right)^{1/2} \left(\frac{R_{*}}{500~R_{\odot}}\right)^{1/6} \nonumber \\ \times \left(\frac{E}{10^{51}~\mathrm{erg}}\right)^{-1/6} \left(\frac{T_{\mathrm{ion}}}{5054~\mathrm{K}}\right)^{-2/3}~\mathrm{days} ,
\end{eqnarray}
where $\kappa$ is the opacity (which, before recombination, is dominated by electron scattering), $M_{\mathrm{ej}}$ is the ejected mass, $R_{*}$ is the radius of the progenitor star, $E$ is the explosion energy, and $T_{\mathrm{ion}}$ is the effective temperature of recombination.
Setting $E=\frac{1}{2}M_{\mathrm{ej}}v_{\mathrm{ej}}^{2}$ to the kinetic energy scaled by the escape velocity, $v_{\mathrm{e}}$, this becomes
\begin{eqnarray}
\label{eqn:maxdur}
t_{\mathrm{p}}\sim 315 \left(\frac{\kappa}{0.34~\mathrm{cm}^{2}\mathrm{g}^{-1}}\right)^{1/6} \left(\frac{M_{\mathrm{ej}}}{10~M_{\odot}}\right)^{1/3} \left(\frac{R_{*}}{500~R_{\odot}}\right)^{1/3} \nonumber \\ \times \left(\frac{T_{\mathrm{ion}}}{5054~\mathrm{K}}\right)^{-2/3} \left(\frac{M_{*}}{25~M_{\odot}}\right)^{-1/6} \left(\frac{v_{\mathrm{ej}}}{v_{\mathrm{e}}}\right)^{-1/3}~\mathrm{days} ,
\end{eqnarray}
where $M_{*}$ is the pre-explosion mass of the progenitor and $v_{\mathrm{ej}}$ is the ejecta velocity.
The ejecta velocity should be larger than $v_{\mathrm{e}}$, but to estimate the maximum duration of a transient powered by recombination we set $v_{\mathrm{ej}}=v_{\mathrm{e}}$ and $M_{\mathrm{ej}}$ to the mass of the hydrogen envelope of the SN progenitor.  

\begin{figure}
\begin{center}
  \includegraphics[width=8.6cm, angle=0]{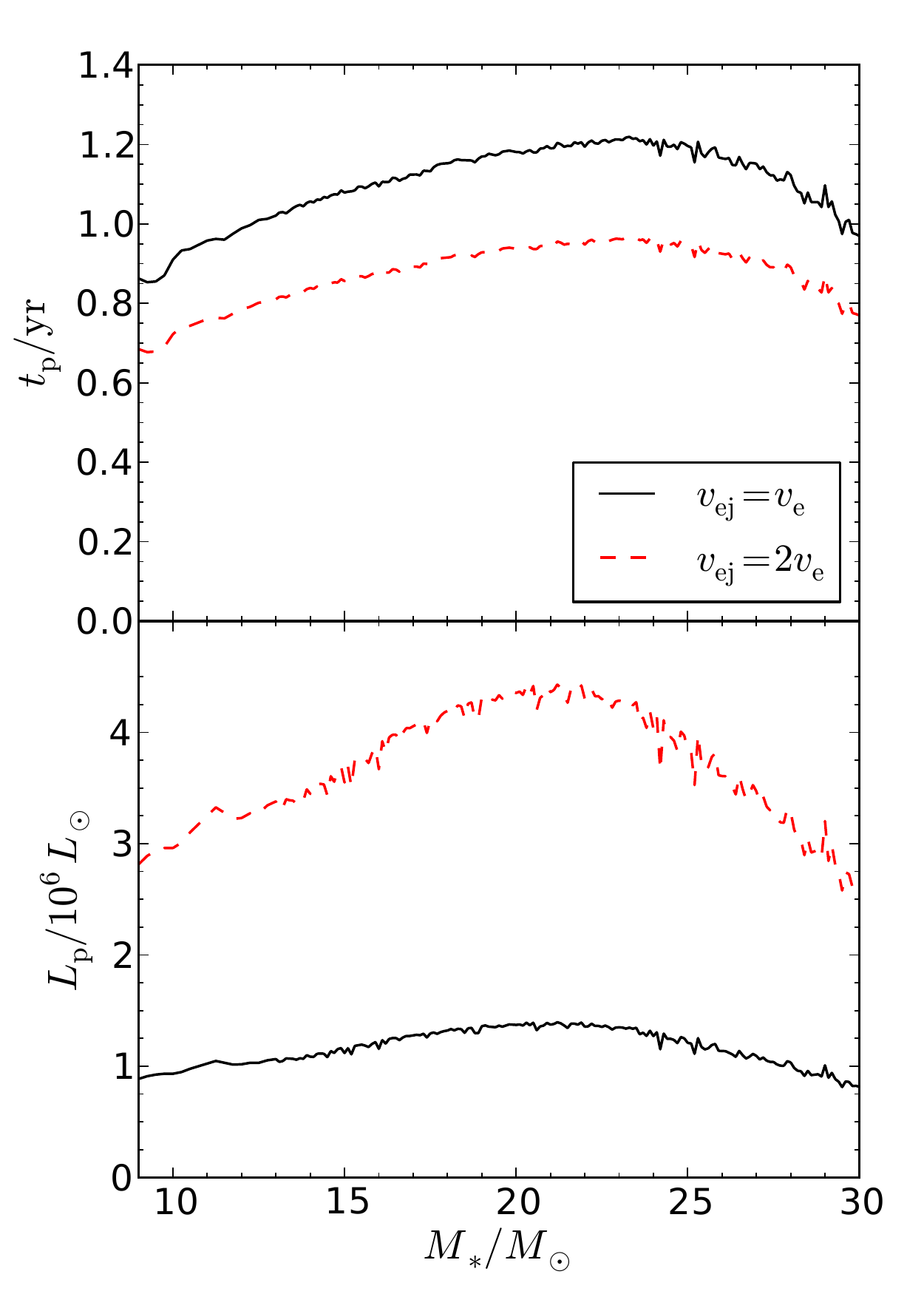}
  \caption{Maximum duration (Eqn. \ref{eqn:maxdur}; top panel) and expected luminosity (Eqn. \ref{eqn:lum}; bottom panel) of a transient powered by recombination for the SN progenitors in \citet{Sukhbold16} for $v_{\mathrm{ej}}=v_{\mathrm{e}}$ (black solid line) and $v_{\mathrm{ej}}=2v_{\mathrm{e}}$ (red dashed line).  Progenitors with initial masses $\gtrsim30~M_{\odot}$ probably do not retain enough of their hydrogen envelopes to power a long-lived transient.  Recombination-powered transients from failed SNe should have luminosities $L\sim10^{6}~L_{\odot}$ that do not last much longer than 1 yr. \label{fig:maxdurlum}}
\end{center}
\end{figure}

Similarly, \citet{Popov93} showed that the luminosity of the recombination-powered plateau phase, $L_{\mathrm{p}}$, is
\begin{eqnarray}
\label{eqn:popovlum}
L_{p} \approx 4.2 \times 10^{8} \left(\frac{\kappa}{0.34~\mathrm{cm}^{2}\mathrm{g}^{-1}}\right)^{-1/3} \left(\frac{M_{\mathrm{ej}}}{10~M_{\odot}}\right)^{-1/2} \nonumber \\ \times \left(\frac{R_{*}}{500~R_{\odot}}\right)^{2/3} \left(\frac{E}{10^{51}~\mathrm{erg}}\right)^{5/6} \left(\frac{T_{\mathrm{ion}}}{5054~\mathrm{K}}\right)^{4/3} ~\mathrm{L_{\odot}} .
\end{eqnarray}
Recasting Eqn. \ref{eqn:popovlum} in terms of the escape velocity, this becomes
\begin{eqnarray}
\label{eqn:lum}
L_{p} \approx 2.3 \times 10^{6} \left(\frac{\kappa}{0.34~\mathrm{cm}^{2}\mathrm{g}^{-1}}\right)^{1/3} \nonumber \\ \times \left(\frac{M_{\mathrm{ej}}}{10~M_{\odot}}\right)^{1/3} \left(\frac{R_{*}}{500~R_{\odot}}\right)^{5/6} \left(\frac{T_{\mathrm{ion}}}{5054~\mathrm{K}}\right)^{4/3} \nonumber \\ \times \left(\frac{M_{*}}{25~M_{\odot}}\right)^{5/6} \left(\frac{v_{\mathrm{ej}}}{v_{\mathrm{e}}}\right)^{5/3}~\mathrm{L_{\odot}} .
\end{eqnarray}
The results using the properties of the SN progenitors in \citet{Sukhbold16} are shown in Fig. \ref{fig:maxdurlum}.  These analytical estimates are roughly consistent with the numerical results of \citet{Lovegrove13} for failed SNe arising from 15 and $25~M_{\odot}$ progenitors.  

Essentially, such a recombination-powered transient should have a luminosity $\sim10^{6}~L_{\odot}$ and not last much longer than $\sim1$ yr.  At late-times the optical emission will also be truncated by dust formation in the dense ejecta \citep{Kochanek14c}, further limiting the optical lifetime.  Thus, we do not consider any of these candidates that have luminosities that decline over multiple years to be promising failed SN candidates (see Table \ref{tab:candidates}).

In addition to these six sources that were selected as failed SN candidates there were an additional 114 sources that slowly faded by $>$$10^{4}~L_{\odot}$ over the entire duration ($\sim7$ yr) of the survey, but were not selected as candidates because an unambiguous point source remained in the final epoch.  We also identified 116 sources that did the reverse --- gradually brightening by $>$$10^{4}~L_{\odot}$ over the entire survey duration.

These slowly fading (and brightening) stars are very rare ($\lesssim10$/galaxy) and no other survey has monitored (with multi-band photometry) so many massive stars over a similar time baseline.
Stars in the Large and Small Magellanic Clouds (and the Galactic bulge) have been monitored in $V$ and $I$ bands for roughly a decade by the OGLE survey.  In the Large Magellanic the closest analogs are two stars that slowly fade $\sim0.3$ mag over $\sim5$ yr and then brighten over the next $\sim5$, and another star that is roughly constant for $\sim6$ years before gradually brightening $\sim1$ mag over the last $\sim4$ yr \citep{Szczygiel10}.  Similarly, \citet{Kourniotis14} present stars in the Small Magellanic Cloud that steadily fade (or brighten) $\sim0.2$ mag over $\sim8$ years.
The slowly fading sources we have identified are notable in that the stars are much more luminous and the fractional change in the luminosity is also much larger.  
In some cases the luminosity changes may be due to increasing optical depths due to increased dust formation, but in many of the cases the sources are blue and are fading roughly achromatically (in $UBVR_{C}$).  Such achromatic changes in the optical could be the result of a changing bolometric correction of a very hot $T_{*}\gtrsim2\times 10^{4}$ K star (e.g., an O star or Wolf-Rayet star).  We note that two of the slowly brightening sources in M101 have already been identified as luminous blue variables \citep{Grammer15}.

\subsection{N925-OC1}
\label{sec:oc}

\begin{figure}
  \includegraphics[width=8.6cm, angle=0]{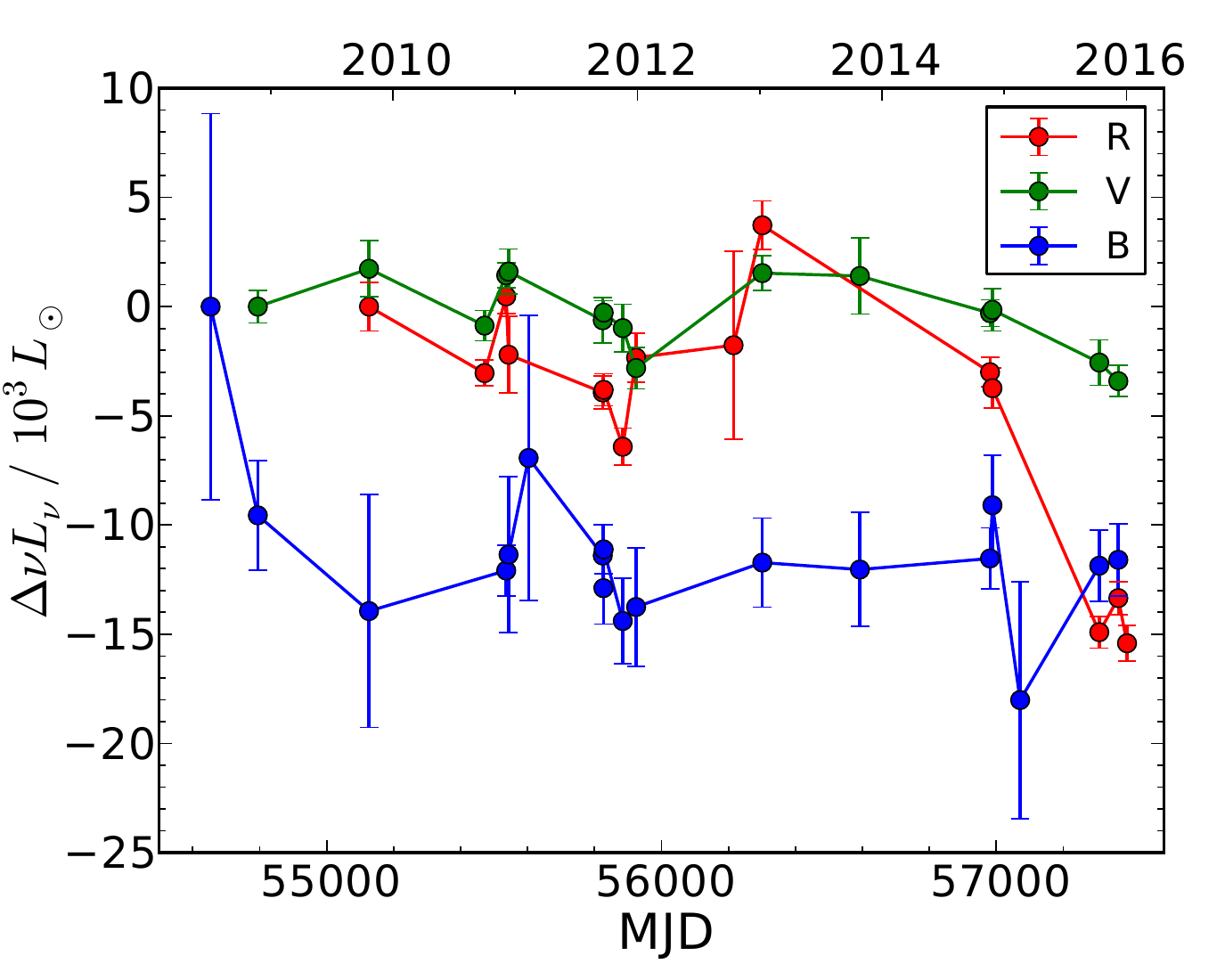}
  \caption{Same as Fig. \ref{fig:PSNlc}, but for N925-OC1. \label{fig:N925-C1lc}}
\end{figure}

\begin{figure*}
\begin{center}
  \includegraphics[width=0.99\textwidth]{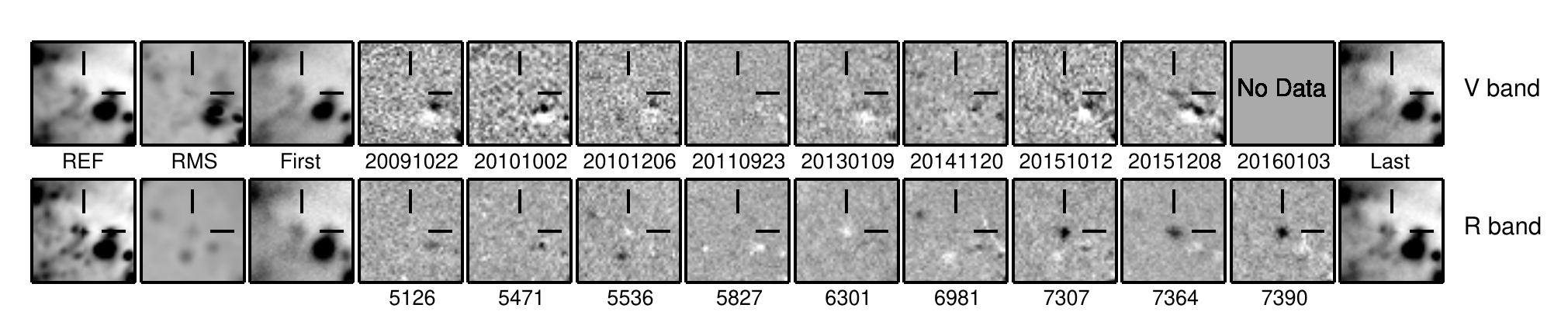}
  \caption{Same as Fig. \ref{fig:PSNstamps}, but for N925-OC1.\label{fig:N925-C1stamps}}
\end{center}
\end{figure*}

The remaining candidate is N925-OC1, which is located at RA 2:27:21.88 and Dec +33:34:05.0.  The $R$ band luminosity dropped by $\sim10^{4}~L_{\odot}$ between successive epochs on 27 November 2014 and 12 October 2015 (Fig. \ref{fig:N925-C1lc}).  The source is in a crowded region and though there is still flux in the final epoch it is unresolved and it is unclear whether it is from a surviving star (Fig. \ref{fig:N925-C1stamps}).
There is no archival \emph{HST} data, but there are 13 \emph{Spitzer Space Telescope} epochs between 13 August 2004 and 14 May 2016 that reveal a flux of $\nu L_{\nu}\sim8\times10^{4}~L_{\odot}$ at $3.6\mu\mathrm{m}$ and $\sim4\times10^{4}~L_{\odot}$ at $4.5\mu\mathrm{m}$ coincident with the progenitor that appears to decrease by $\sim10\%$ when the $R$ band luminosity dropped.
The observed decline in optical and IR flux is consistent with the disappearance of a $L_{*}\sim10^{4.7}~L_{\odot}$ star with $T_{\mathrm{eff}}\sim3500$ K and a modest amount of circumstellar dust ($\tau_{V}\sim0.4$), which would correspond to an initial mass of $\sim10~M_{\odot}$.  However, we note that this source has been at its observed minimum for only 3 epochs over 3 months, increasing the likelihood that it could be a false positive.  Given that the survey has identified several similar sources in crowded fields where a surviving star is undetected with the LBT's resolution but is clearly resolved with \emph{HST} imaging, we presently consider this source a likely variable rather star than a failed SN.
  Additional monitoring with the LBT is needed to further vet this candidate before pursuing more expensive follow-up with space-based observations.


\section{Rate of Failed Supernova}
\label{sec:rate}

\begin{figure}
  \includegraphics[width=8.6cm, angle=0]{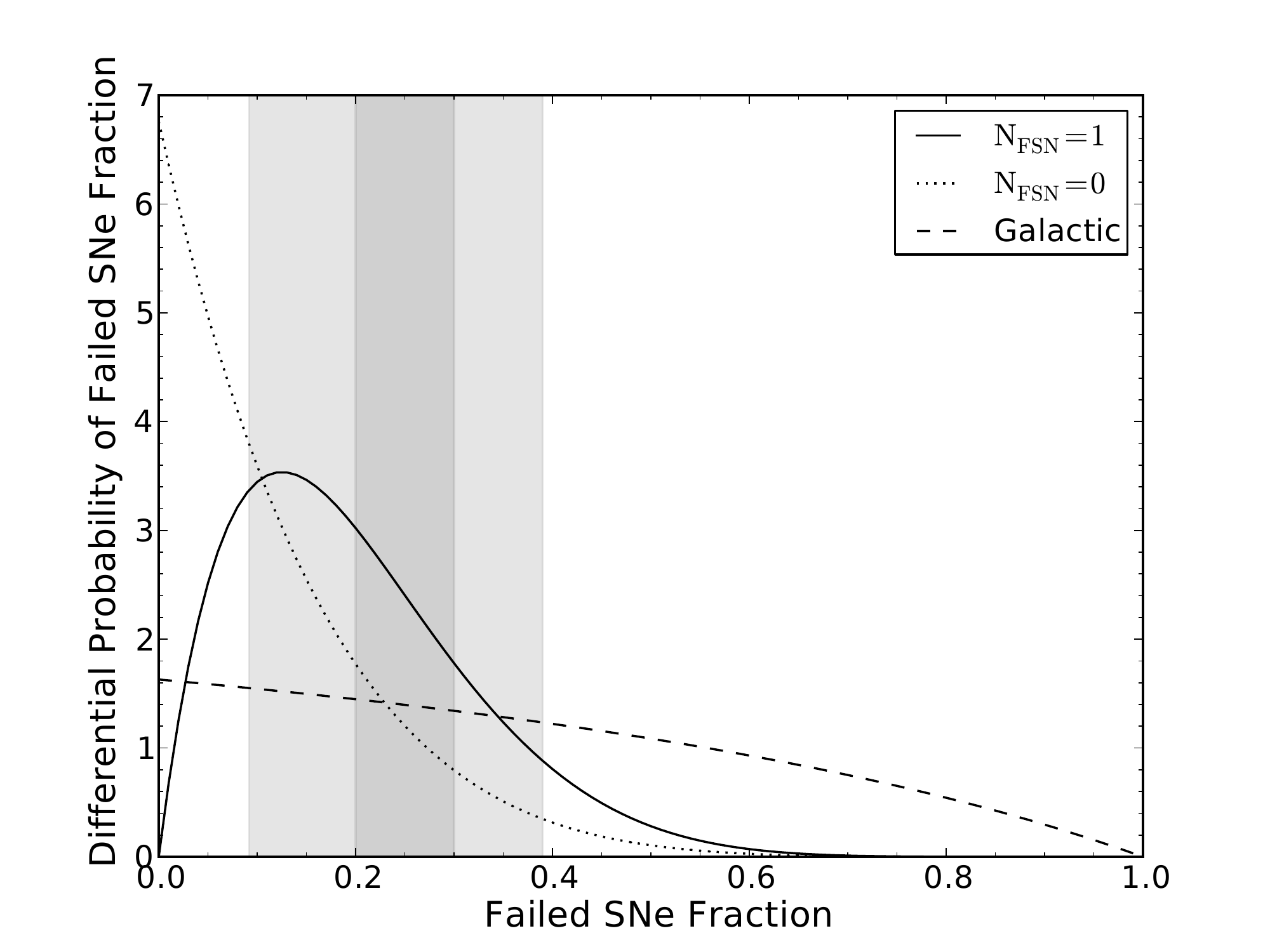}
  \caption{Probability for the fraction of core-collapses that become failed SNe if there are $\mathrm{N}_{\mathrm{FSN}}=0$ (dotted) or $\mathrm{N}_{\mathrm{FSN}}=1$ (solid) failed SNe in our sample.  The dashed line shows the constraints from combining the historical Galactic SN rate with the non-detections of neutrinos from a Galactic ccSN \citep{Adams13}.  The black hole mass function suggests $0.092<f<0.39$ \citep{Kochanek15} as shown by the lightly shaded region, and the RSG problem is solved if $f\sim0.2-0.3$ as shown by the darkly shaded region. \label{fig:prob}}
\end{figure}

We view N6946-BH1 as the only probable failed SN identified thus far by the survey.  During the selection window there have been $N_{\mathrm{SN}}=6$ successful ccSNe.
If we assume the survey is complete for both successful and failed SNe, then the fraction of core-collapses resulting in failed SNe, $f$, is simply given by the binomial distribution
\begin{equation}
P(f) \propto (1-f)^{N_{\mathrm{SN}}} f^{N_{\mathrm{FSN}}} ,
\end{equation}
where $N_{\mathrm{FSN}}$ is the number of failed ccSNe within the survey and the normalization is set by $\int^{1}_{0} P(f) \mathrm{d}f \equiv 1$.  For $N_{\mathrm{FSN}}=1$ and $N_{\mathrm{SN}}=6$ the median fraction of failed core collapses is 0.143 with a $90\%$ confidence interval of $0.046 \leq f \leq 0.471$ (see Fig. \ref{fig:prob}).

However, the survey's completeness to failed SNe is naturally lower than the completeness for successful SNe.  For example, the Type IIP SN 2014bc was in the saturated core of NGC 4258 and likely would have been missed by our survey if the core-collapse had instead been a failed SN.  
As already noted in GKS15, we will be incomplete for failed SNe with optical ($UBVR_{\mathrm{C}}$) luminosities $\leq 10^{4}~L_{\odot}$, which will exclude the lowest mass systems to undergo core collapse ($M\lesssim9~M_{\odot}$) and hot, stripped progenitors ($M\gtrsim30~M_{\odot}$).  The survey is, however, very sensitive to the more massive RSGs that are most likely associated with failed SNe.

The main goal of the survey is to answer the question of the missing high-mass RSG progenitors.  For these progenitors the main causes of incompleteness are likely events occurring near the cores of galaxies where our $R$ band imaging is saturated and possibly failed SNe occurring within compact, unresolved clusters that we mistake for a surviving star.  Out of the 6 SNe in the sample, one occurred within a luminous cluster \citep[SN 2012fh;][]{Fraser12}.  Because this was a SN Ic with an optically faint progenitor we cannot use the event to directly test whether a failed SN from RSG at this position would have been selected as a candidate.  Thus, though we have chosen to present our failed SN fraction based on observing 6 luminous SNe in the sample, it arguably could be based relative to 4 or 5 SNe to account for incompleteness to failed SNe.  At this point in the survey, however, Poisson fluctuations in the numerator are far more important than the exact completeness fraction.

If N6946-BH1 is ultimately rejected as a failed SN, the upper limit on the failed SN fraction is $f < 0.35$ at $90\%$ confidence.  
These constraints are a significant improvement over the earlier ones reported by GKS15 and are the best direct measurements of the failed SN fraction.  For comparison, the combination of the Galactic SN rate with the absence of any neutrino detections \citep{Alexeyev02,Ikeda07} of a failed SN over the last $\sim$3 decades yields the much weaker constraint that $f\leq0.69$ (at $90\%$ confidence; \citealt{Adams13}).  Indirect constraints from the missing RSG progenitors suggest that $f\sim0.2-0.3$ and the black hole mass function can be explained by $0.09\leq f \leq0.39$ \citep{Kochanek15}.  Our direct constraints are consistent with these estimates.

The LBT survey is expected to continue for at least 3 more years (for a minimum total duration of 10 yr) and there has already been one additional SN in NGC 3627 \citep[ASAS-SN16fq;][]{Bock16,Kochanek17} since the end of the candidate selection window.  Based on both historical SN observations and the SNe that have occurred during the survey, the SN rate of the galaxies we are monitoring is $\sim1/\mathrm{yr}$ (GKS15).  Accordingly, the expectation by the end of the survey will be to have the failed SN fraction constrained within $\sim15\%$ and know whether failed SN explain the missing RSG SN progenitors and the black hole mass function.  

\section*{Acknowledgements}
We thank Adam Leroy for early access to his proprietary \emph{HST} imaging.
Financial support for this work was provided by NSF through grant AST-1515876.
This work is based on observations made with the Large
Binocular Telescope. The LBT is an international collaboration
among institutions in the United States, Italy, and Germany. The
LBT Corporation partners are: the University of Arizona on behalf
of the Arizona university system; the Istituto Nazionale di
Astrofisica, Italy; the LBT Beteiligungsgesellschaft, Germany,
representing the Max Planck Society, theAstrophysical Institute
Potsdam, and Heidelberg University; the Ohio State University;
and the Research Corporation, on behalf of the University of
Notre Dame, the University of Minnesota, and the University of
Virginia.
This work also utilized observations made with the NASA/ESA Hubble Space Telescope, obtained from the data archive at the Space Telescope Science Institute. STScI is operated by the Association of Universities for Research in Astronomy, Inc. under NASA contract NAS 5-26555.

\bibliography{references}
\bibliographystyle{mn2e}
\clearpage
\end{document}